\documentclass[preprint,12pt]{elsarticle}

\usepackage{amssymb}
\usepackage{graphicx}
\usepackage{dcolumn}

\newcommand{\lb}{\left(}
\newcommand{\rb}{\right)}

\newcommand{\Lb}{\left\{}
\newcommand{\Rb}{\right\}}
\newcommand{\beq}{\begin{equation}}
\newcommand{\eeq}{\end{equation}}
\newcommand{\beqn}{\begin{eqnarray}}
\newcommand{\eeqn}{\end{eqnarray}}
\journal{Physics Letter B}
\begin{document}

\begin{frontmatter}

\title{Chirality in odd-$A$ nucleus $^{135}$Nd in particle rotor model}

\author[PKU]{B. Qi}

\author[PKU,ITP]{S.Q. Zhang\corref{corr1}}\ead{sqzhang@pku.edu.cn}

\author[PKU,ITP,LZ,Stell]{J. Meng\corref{corr1} } \ead{mengj@pku.edu.cn}

\author[SDU]{S.Y. Wang}

\author[USA]{S. Frauendorf }

\cortext[corr1]{Corresponding author}
\address[PKU]{State Key Lab Nucl. Phys. {\rm\&} Tech., School of Physics, Peking University, Beijing 100871, China}
\address[ITP]{Institute of Theoretical Physics, Chinese Academy of Science, Beijing 100080, China}
\address[LZ]{Center of Theoretical Nuclear Physics, National Laboratory of Heavy Ion Accelerator, Lanzhou 730000, China}
\address[Stell]{Department of Physics, University of Stellenbosch, Stellenbosch, South Africa}
\address[SDU]{Department of Space Science and Applied Physics, Shandong University at Weihai, Weihai 264209, China}
\address[USA]{Physics Department, University of Notre Dame, Notre Dame, Indiana 46556, USA}

\begin{abstract}
A particle rotor model is developed which couples several valence
protons and neutrons  to a rigid triaxial rotor core. It is applied
to investigating the chirality in odd-$A$ nucleus $^{135}$Nd with
$\pi h_{11/2}^2\otimes\nu h^{-1}_{11/2}$ configuration for the first
time in a fully quantal approach. For the two chiral sister bands,
the observed energies and the $B(M1)$ and $B(E2)$ values for the
in-band  as well as interband transitions are reproduced
excellently. Root mean square values of the angular momentum
components and their probability distributions are used for
discussing in detail the chiral geometry of the aplanar rotation and
its evolution with angular momentum. Chirality is found to change
from a soft chiral vibration to nearly static chirality at spin
$I=39/2$ and back to another type of chiral vibration at higher
spin.
\end{abstract}


\end{frontmatter}


Spontaneous chiral symmetry breaking is a phenomenon of general
interest in chemistry, biology and particle physics. Since the
pioneering work of nuclear chirality in 1997~\cite{FM97}, much
effort has been devoted to further explore this interesting
phenomenon. Following the observation of chiral doublet bands in
$N=75$ isotones~\cite{Starosta01}, more candidates have been
reported over more than 20 nuclei experimentally in A$\sim100$, 130
and 190 mass regions including odd-odd, odd-$A$ and even-even nuclei
~\cite{Koike01,Bark01,Mergel02, Koike03, Zhu03, Joshi04,
Alcantara04, Timar04, Vaman04, Timar06, WangSY06a, Lawrie08}.

Chiral symmetry breaking was initially suggested to occur in a
stable triaxial deformed nucleus, with a high-$j$ particle-like
valence proton (neutron) and a high-$j$ hole-like valence neutron
(proton)~\cite{FM97}. In this case, the angular momenta of the core,
the valence proton and neutron are mutually perpendicular, arranging
in the body-fixed frame into two systems with opposite chirality
(left- and right-handed). Due to the restoration of chiral symmetry
by quantum tunneling, a pair of near degenerate $\Delta I = 1$ bands
-the chiral sister or doublet bands- are
observed~\cite{FM97,Frauendorf01}.

Theory wise, chiral doublet bands were first investigated in the
one-particle-one-hole-rotor model (PRM) and the corresponding tilted
axis cranking (TAC) approximation~\cite{FM97}. Later on realistic
TAC approaches, as the Strutinsky shell correction method with a
hybrid Woods-Saxon and Nilsson potential~\cite{Dimitrov00PRL}, the
Skyrme Hartree-Fock model~\cite{Olbratowski04}, as well as the
relativistic mean field model~\cite{MengJ06,PengJ08} has been
developed to investigate this new phenomena. Within the TAC mean
field approximation, the left-handed and right-handed solutions are
exactly degenerate. It is not possible to calculate the energy
difference between the bands, which is the consequence of quantum
tunneling between the two solutions. Before the onset of chirality
of the mean field, the precursor of the symmetry breaking occurs as
a soft vibration between the right- and left-handed configurations.
These chiral vibrations have been studied in the framework of the
random phase approximation (RPA) based on the TAC mean field
~\cite{Mukhopadhyay07,Almehed08}. However, TAC+RPA is not able to
describe the smooth transition from a slow vibration to quantum
tunneling between the left- and right-handed mean field solutions.
Moreover, angular momentum is not a good quantum number and the
electromagnetic transitions are calculated in semiclassical
approximation.

The fully quantal PRM does not suffer from these deficiencies. Total
angular momentum is a good quantum number. The energies and
transition probabilities are treated fully quantal. In particular,
it provides the energy splitting between doublet bands covering the
whole range from chiral vibrations to weak tunneling between the two
chiral configurations. On the other hand, the PRM is based the
assumption that the rotor has a fixed deformation. This may appear
as a problem, because the nuclei in which chiral doublets have been
found are considered to be soft with respect to the triaxiality
parameter $\gamma$. The study in the framework of an IBA core
coupled to a particle and a hole found a substantial coupling
between the shape and angular momentum orientation degrees of
freedom~\cite{Tonev06}. In contrast, the microscopic studies in the
framework of TAC+RPA~\cite{Mukhopadhyay07,Almehed08} indicate that
the angular momentum dynamics is almost decoupled from from the
shape degrees of freedom, which justifies the application of the
PRM.

For odd-odd nuclei in $A\sim100$ and $130$ regions, chirality has
been extensively studied with PRM with $1$-particle-$1$-hole
configuration~\cite{PengJ03,Koike04}, or by introducing pairing to
simulate the effect of many valence nucleons~\cite{Koike03,
WangSY07,Zhang07,WangSY08}. Good agreement with the experimental
spectra and electromagnetic transitions has been obtained by PRM for
the doublet bands in odd-odd nuclei $^{126}$Cs~\cite{WangSY07},
$^{128}$Cs~\cite{Koike03, Grodner06}, and
$^{106}$Rh~\cite{WangSY08}.

Apart from odd-odd nuclei, chiral doublet bands have been observed
in odd-$A$ nuclei. The first example is the chiral spectra
characteristic observed in $^{135}$Nd~\cite{Zhu03}, which is further
confirmed by the lifetime measurements~\cite{Mukhopadhyay07} and
suggested to be built on the configuration $\pi
h^{2}_{11/2}\otimes\nu h^{-1}_{11/2}$. More candidates have been
reported in $^{103}$Rh~\cite{Timar06},
$^{105}$Rh~\cite{Alcantara04,Timar04} as well. For even-even nuclei,
a candidate was suggested for $^{136}$Nd~\cite{Mergel02}, however
the interpretation is doubted on the basis of the recent lifetime
measurement~\cite{Mukhopadhyay08}. The existence of these data makes
a PRM that treats more than one valence proton and one valence
neutron  highly desirable. A PRM with an axial or a triaxial core
coupled to many-particle configurations has been developed and used
for the description of magnetic bands of
$^{198,199}$Pb~\cite{Carlsson06} and the wobbling excitation in
$^{163}$Lu~\cite{Carlsson07}. However the study of chirality with
such kind of model is still absent. There has been a dispute on the
identification and interpretation of chiral doublet bands based on
$1$-particle-$1$-hole configuration in odd-odd
nuclei~\cite{Tonev06,Grodner06,Mukhopadhyay08,Petrache06,Joshi07}.
The study of the doublet bands in odd-A nuclei in the framework of a
rotor coupled to two particles and one hole should shed new light on
the question of nuclear chirality.

In this Letter, a triaxial $n$-particle-$n$-hole PRM will be
developed to treat more than one valence proton and one valence
neutron and applied to the study of nuclear chirality. The energy
spectra and electromagnetic transitions probabilities of the doublet
bands in $^{135}$Nd will be calculated and compared with the data
available as well as the previous TAC+RPA results. The structure of
doublet bands with $2$-particle-$1$-hole configuration will be
discussed.

The total Hamiltonian is expressed as,
 \beq\label{eq:multiPRM}
  \hat H= \hat H_\textrm{coll}+ \hat H_\textrm{intr},
 \eeq
with the collective rotor Hamiltonian $H_\textrm{coll}$,
 \beq
 \hat H_\textrm{coll}=\sum_{k=1}^3
 \frac{\hat R_k^2}{2{\cal{J}}_k}=\sum_{k=1}^3
 \frac{(\hat I_k-\hat J_k)^2}{2{\cal{J}}_k},
 \eeq
where the indices $k=1,2,3$ refer to the three principal axes of the
body-fixed frame, $\hat{R}_k, \hat{I}_{k}, \hat{J}_{k}$ denote the
angular momentum operators for the core, the total nucleus and the
valence nucleons, respectively. The moments of inertia for
irrotational flow are adopted, i.e., ${\cal J}_k = {\cal
J}_0\sin^2(\gamma - {2\pi k}/{3})$. The intrinsic Hamiltonian for
valence nucleons is
 \beq \label{eq:sp}
 \hat H_\textrm{intr}
  = \sum_{\nu}\varepsilon_{p,\nu}
  a_{p,\nu}^{+}a_{p,\nu}
  + \sum_{\nu'}\varepsilon_{n,\nu'}
  a_{n,\nu'}^{+}a_{n,\nu'},
 \eeq
where $\varepsilon_{p,\nu}$ and  $\varepsilon_{n,\nu'}$ are the
single particle energy of proton and neutron.

The single particle states are expressed as
 \beq \label{eq:spwf}
 {a}^{+}_{\nu}| 0 \rangle
 =\sum_{\alpha \Omega}c_{\alpha \Omega}^{(\nu)}
 |\alpha,\Omega \rangle,
    ~~~~
 {a}^{+}_{\overline{\nu}}| 0 \rangle
 = \sum_{\alpha \Omega}(-1)^{j-\Omega}c_{\alpha \Omega }^{(\nu)}
 |\alpha,-\Omega \rangle,
 \eeq
where $\Omega$ is the projection of the single-particle angular
momentum ${\hat j}$ along the 3-axis and is restricted to half of
the values, $-j, \cdots, -1/2, 1/2, 3/2, \cdots, j$, due to the
time-reversal degeneracy~\cite{Larsson78}, and $\alpha$ denotes the
other quantum numbers. For a system with $z$ valence protons and $n$
valence neutrons, the intrinsic wave function is given as
 \beq\label{eq:configurations}
 |\varphi\rangle = \lb\prod_{i=1}^{z_{1}}a^\dag_{p, \nu_i}\rb
 \lb\prod_{i=1}^{z_{2}}a^\dag_{p,\overline{\mu_i}}\rb
 \lb\prod_{i=1}^{n_{1}}a^\dag_{n,\nu'_i}\rb
 \lb\prod_{i=1}^{n_2}a^\dag_{n,\overline{\mu'_i}}\rb
 | 0\rangle
 \eeq
with $z_{1}+z_{2}=z, n_1+n_2=n$, $0\leq z_{1}\leq z, 0\leq n_1\leq
n$.

The total wave function can be expanded into the strong coupling
basis,
 \beq
 |IM\rangle = \sum_{K\varphi}c_{K\varphi}|IMK\varphi\rangle,\label{eq:expansion}
 \eeq
with
 \beq
 |IMK\varphi\rangle
   =  \frac{1}{\sqrt{2(1+\delta_{K0}\delta_{\varphi,\overline{\varphi}})}}
  \lb |IMK\rangle |\varphi\rangle
  +(-1)^{I-K} |IM -K\rangle|\overline{\varphi}\rangle\rb, \label{eq:basis}
 \eeq
where $|IMK\rangle$ denotes the Wigner functions
$\sqrt{\frac{2I+1}{8\pi^2}}D^I_{MK}$ and $\varphi$ is a shorthand
notation for the configurations in Eq.~(\ref{eq:configurations}).
The basis states are symmetrized under the point group $D_2$, which
leads to $K-\frac{1}{2}(z_1-z_2)-\frac{1}{2}(n_1-n_2)$ being an even
integer with $\Omega=\cdots, -3/2, 1/2, 5/2, \cdots$. The reduced
transition probabilities $B(M1)$ and $B(E2)$ can be obtained from
the wave function of PRM with the $M1$ and $E2$
operators~\cite{Zhang07}.

The single particle energy for proton and neutron
$\varepsilon_{p,\nu}$ and  $\varepsilon_{n,\nu'}$ in
Eq.~(\ref{eq:sp}) are provided by the triaxial deformed single-$j$
shell Hamiltonian~\cite{FM97},
 \beq\label{eq:hsp}
  h_\textrm{sp}=\pm \frac{1}{2}C
  \Lb\cos\gamma(\hat j_3^2-\frac{j(j+1)}{3})
  + \frac{\sin\gamma}{2\sqrt{3}}(\hat j_+^2+\hat j_-^2)\Rb,
 \eeq
where the plus or minus sign refers to particle or hole, and the
coefficient $C$ is proportional to the quadrupole deformation
$\beta$ as in Ref.~\cite{WangSY08}.

In the PRM calculations for the doublet bands in $^{135}$Nd, the
configuration $\pi h^{2}_{11/2}\otimes\nu
h^{-1}_{11/2}$~\cite{Mukhopadhyay07} is adopted. The deformation
parameters $\beta=0.235$ and $\gamma=22.4^{\circ}$ for $^{135}$Nd
are obtained from the microscopic self-consistent triaxial
relativistic mean field calculation~\cite{MengJ06}. Accordingly,
$C_p$ and $C_n$ in Eq.~(\ref{eq:hsp}) take values of $0.323$ and
$-0.323$ MeV~\cite{WangSY08}. The moment of inertia ${\cal
J}_0=29.0$ MeV$/\hbar^2$ is adjusted to the experimental energy
spectra. For the electromagnetic transition, the empirical intrinsic
quadrupole moment $Q_0=(3/\sqrt{5\pi})R_0^2Z\beta=4.0$ eb,
gyromagnetic ratios $g_R=Z/A=0.44$, and $g_p$=1.21, $g_n$=-0.21 are
adopted~\cite{WangSY07}.

\begin{figure}[h!]
 \centering
 \includegraphics[width=13cm]{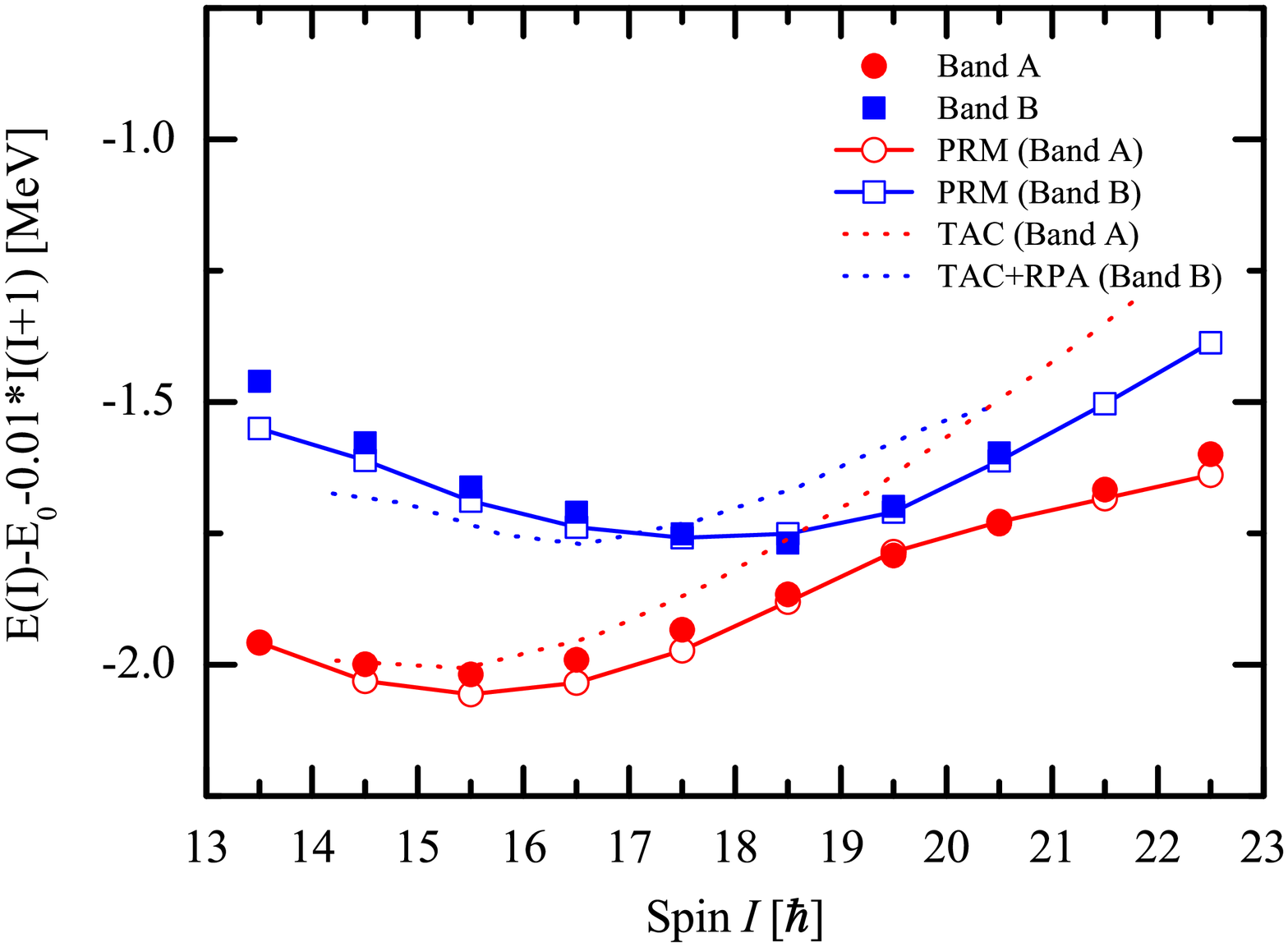}
 \caption{(color online) The excitation energies $E(I)$ for the
chiral sister bands in $^{135}$Nd calculated by means of the
triaxial PRM with configuration $\pi h_{11/2}^2\otimes\nu
h^{-1}_{11/2}$ (open symbols) in comparison with the data (filled
symbols)~\cite{Zhu03,Mukhopadhyay07} and the corresponding TAC+RPA
results (dotted lines)~\cite{Mukhopadhyay07}. The energies are
relative to the band head $E_0$ of the chiral bands, with a rotor
reference subtracted. } \label{fig:energy}
\end{figure}

The calculated excitation energy spectra $E(I)$ for the doublet
bands A and B in $^{135}$Nd are presented in Fig.~\ref{fig:energy},
together with the corresponding data~\cite{Zhu03, Mukhopadhyay07}.
The experimental energy spectra are excellently reproduced by the
PRM calculation.  Except for the 27/2-state in band B, the
calculated results agree with the data within 50 keV. In particular,
the trend and amplitude for the energy splitting between two partner
bands are excellently reproduced.

For comparison, the TAC+RPA results~\cite{Mukhopadhyay07} are also
included. The energy splitting between two sister bands is well
described for  $I \leq 39/2$, which is the vibrational regime, where
the RPA is stable. The energy splitting disappears at $I=41/2$,
where the TAC solution attains chirality. The description of the
experiment in the region $I>39/2$ is beyond the realm of RPA based
on the planar TAC solution. Being a quantum theory that is not
restricted to small amplitude vibrations, PRM is able to perfectly
reproduce the energy splitting for the whole observed spin region,
staying within the configuration $\pi h^{2}_{11/2}\otimes\nu
h^{-1}_{11/2}$ and the triaxial rotor. The analysis below shows that
the motion of the angular momentum vector in fact changes from
chiral vibration about the plane spanned by the long and short axes
of the triaxial shape to tunneling between the left- and
right-handed configurations.

\begin{figure}[h!]
  \centering
  \includegraphics[height=13cm]{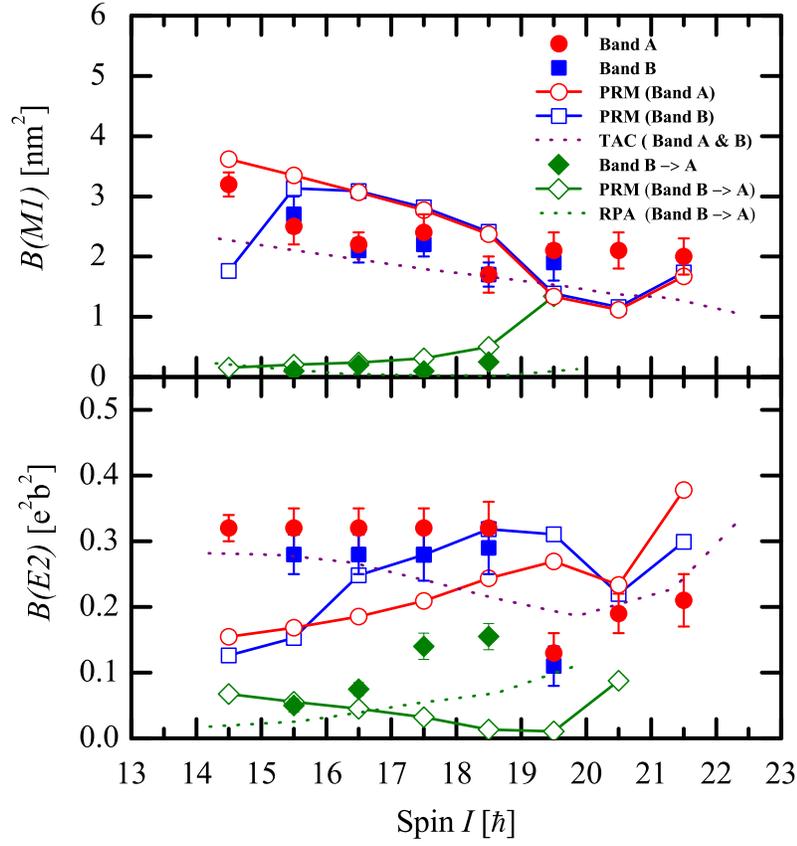}
  \caption{(color online) The  $B(M1)$ and $B(E2)$ values calculated
by means of the PRM  for the chiral sister bands in $^{135}$Nd (open
symbols) in comparison with the data (filled
symbols)~\cite{Mukhopadhyay07} and the corresponding TAC+RPA results
(dotted lines)~\cite{Mukhopadhyay07}. } \label{fig:M1E2}
\end{figure}

The calculated in-band and interband transition probabilities
$B(M1)$ (upper panel) and $B(E2)$ (lower panel) for the doublet
bands in $^{135}$Nd are presented in Fig.~\ref{fig:M1E2}, together
with the available data and the TAC+RPA
results~\cite{Mukhopadhyay07}. For $I \leq 39/2$, the observed
in-band $B(M1)$ values in the two bands are almost the same and much
larger than the interband ones. These features are perfectly
reproduced by the PRM calculation. The TAC+RPA gives a similar
result. After $I=39/2$ the two models differ. In the case of PRM,
the interband and in-band $B(M1)$ transitions become comparable due
to the transition to tunneling regime, which is absent in the
TAC+RPA calculations.

For $I \leq 39/2$, the observed in-band $B(E2)$ values, in
particular their similarity, are well reproduced by both the PRM and
TAC+RPA calculations, while the observed interband $B(E2)$ values
are underestimated in both the PRM and TAC calculations. The kink at
$I=39/2$ of the in-band $B(E2)$ values of the PRM reflects the
transition to tunneling regime as mentioned in the discussion of
$B(M1)$. The ratio between the in-band and interband $B(E2)$ and
$B(M1)$ values depends sensitively on the details of the transition
from the vibrational to the tunneling regime, which may account for
the deviations of the PRM calculation from experiment.

\begin{figure}[h!]
  \centering
  \includegraphics[width=13cm, bb=0 30 720 650]{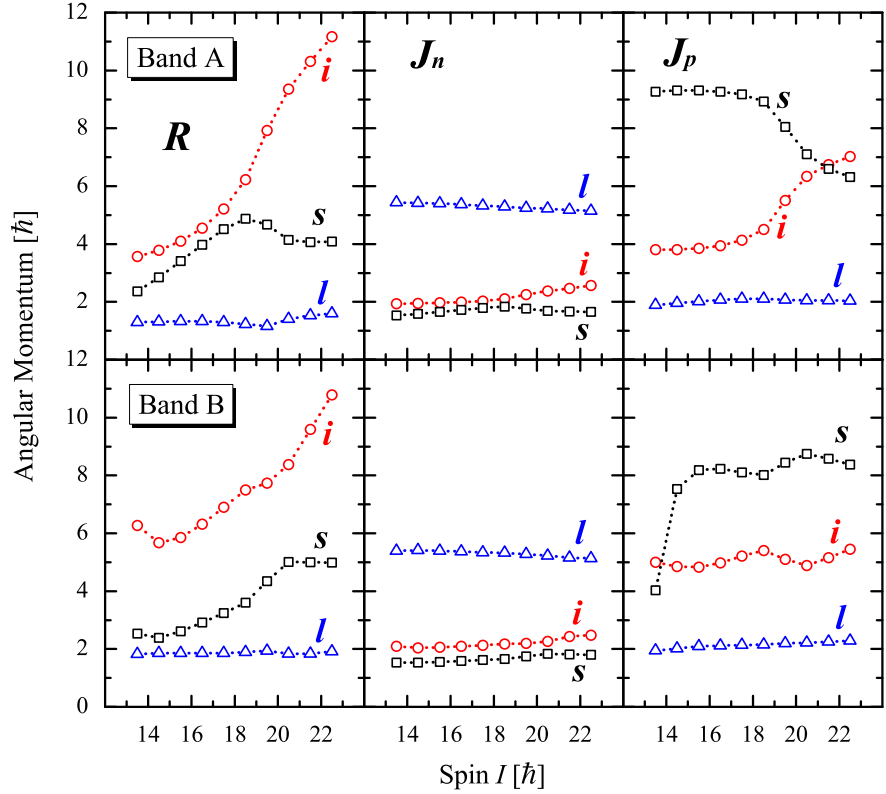}
\caption{(color online) The root mean square components along the
intermediate ($i$-, circles), short ($s$-, squares) and long ($l$-,
triangles) axis of the core ${R_k} = \sqrt{\langle \hat{R}_{k}^2
\rangle}$, valence neutron $J_{nk} = \sqrt{\langle \hat{j}_{nk}^2
\rangle}$, and valence protons angular momenta $J_{pk}=\sqrt{\langle
(\hat{j}_{(p1)k}+\hat{j}_{(p2)k})^2 \rangle}$ calculated as
functions of spin $I$ by means of the PRM for the doublet bands in
$^{135}$Nd. } \label{fig:spin1}
\end{figure}

\begin{figure}[h!]
  \centering
  \includegraphics[width=6cm, bb=150 130  540 400]{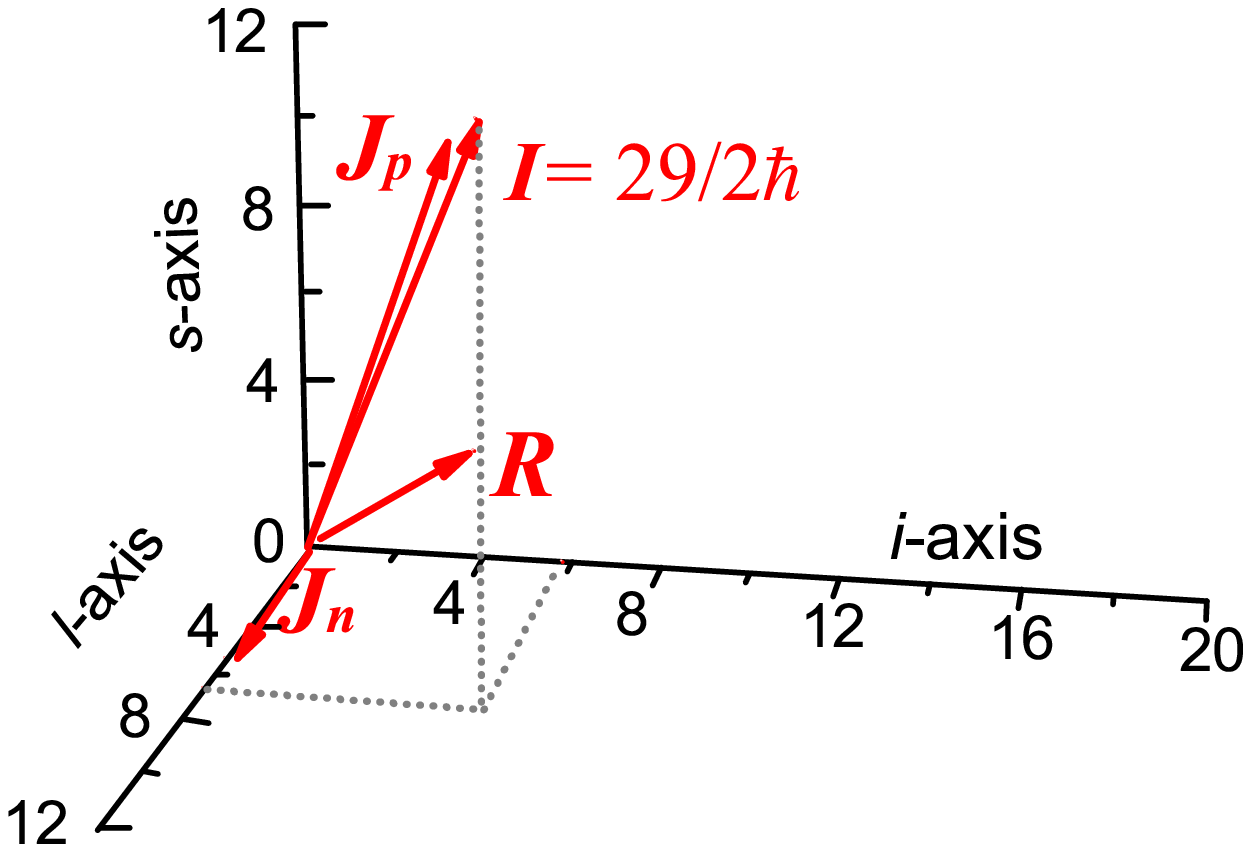}
  \includegraphics[width=6cm, bb=150 130  540 400]{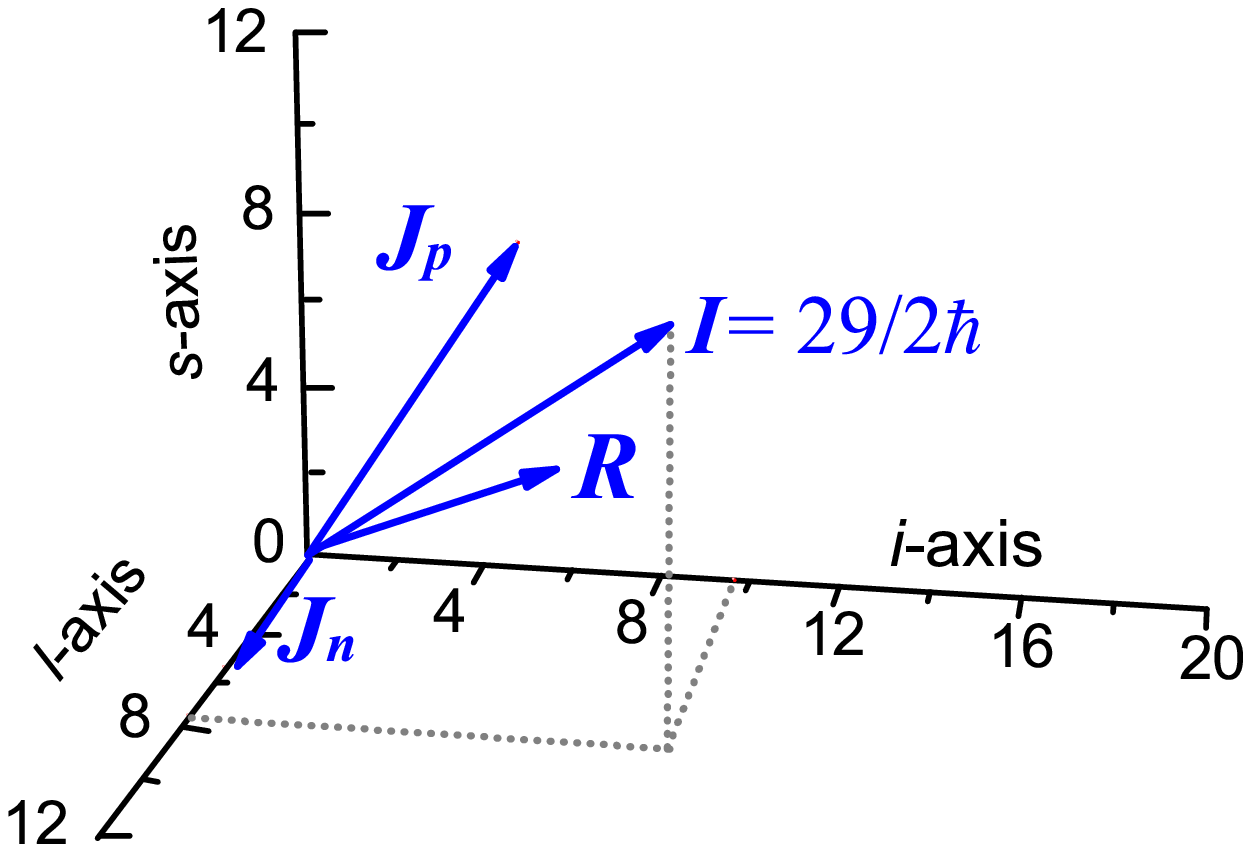}
  \includegraphics[width=6cm, bb=150 130  540 400]{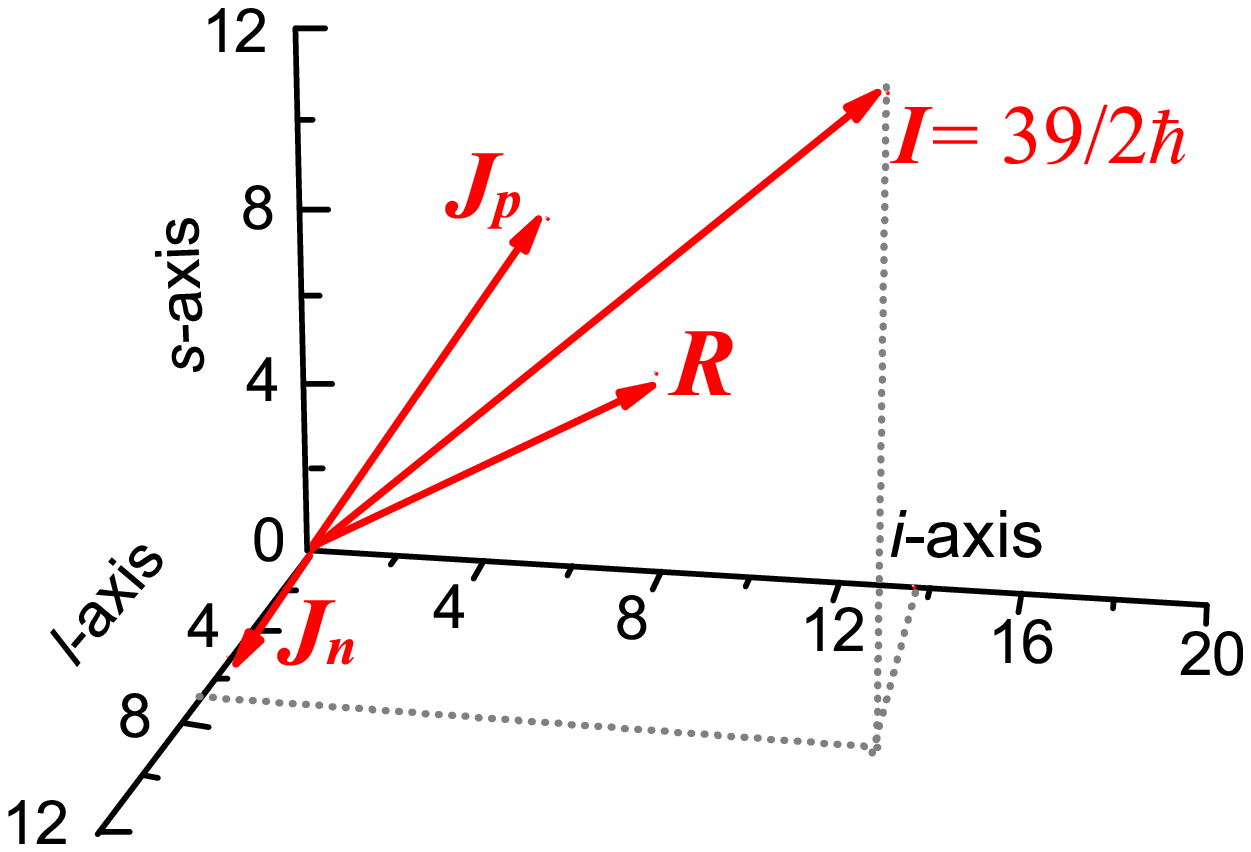}
  \includegraphics[width=6cm, bb=150 130  540 400]{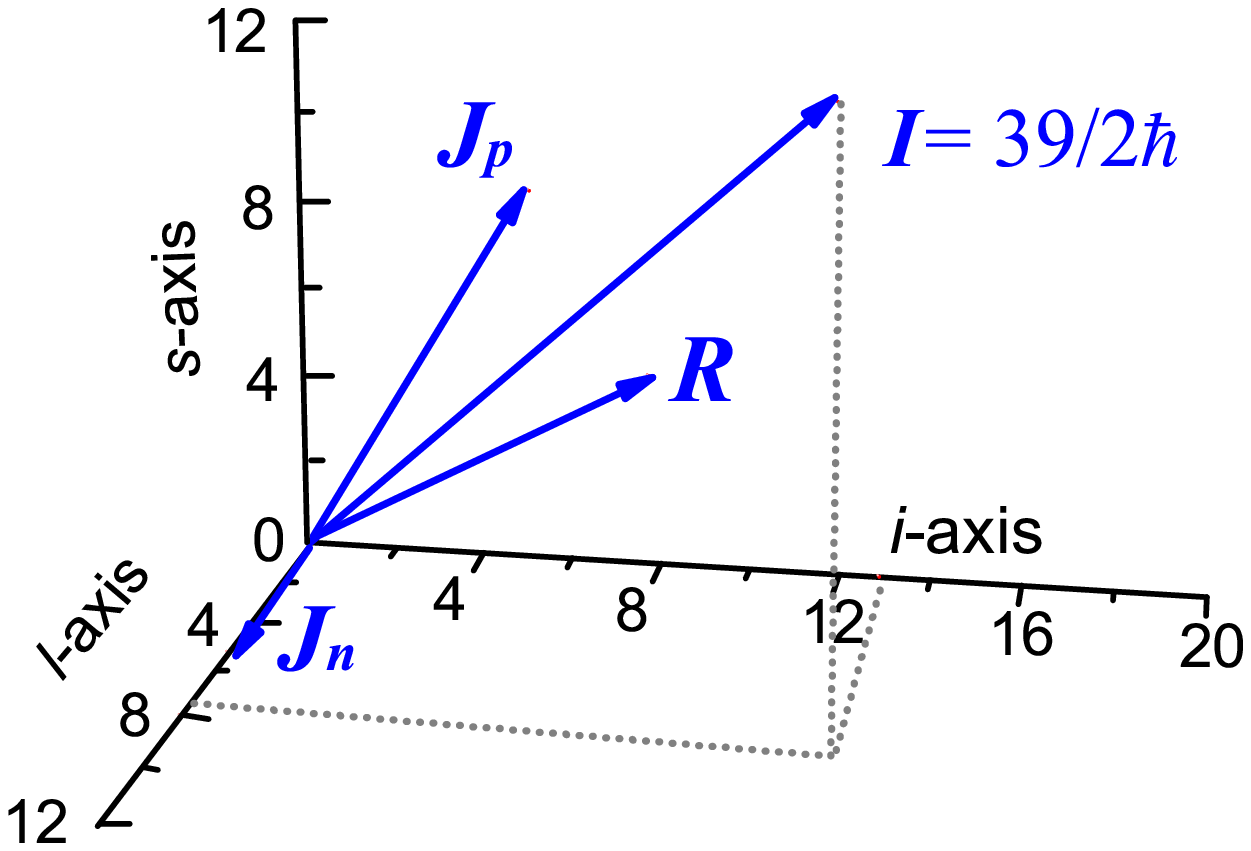}
  \includegraphics[width=6cm, bb=150 100  540 400]{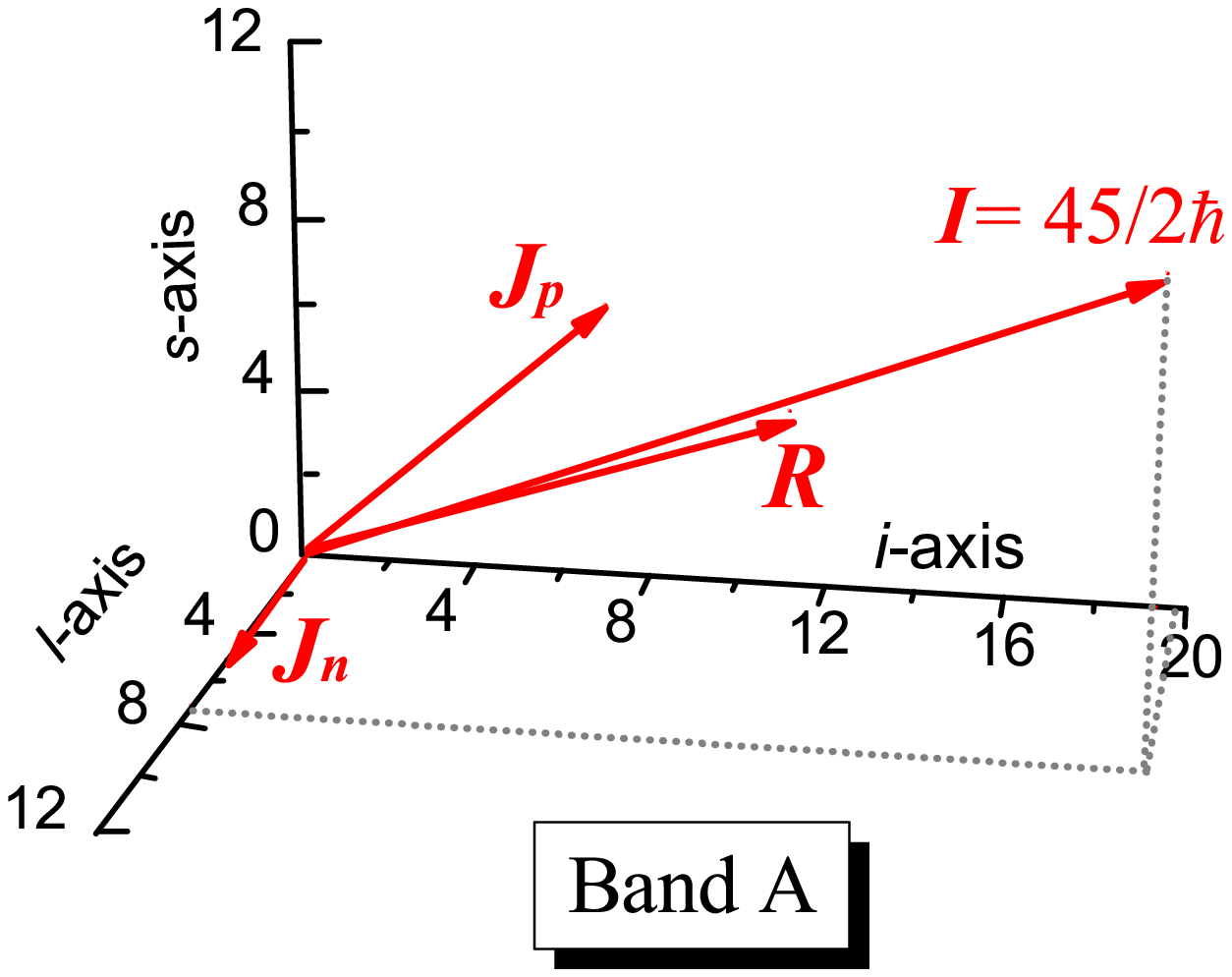}
  \includegraphics[width=6cm, bb=150 100  540 400]{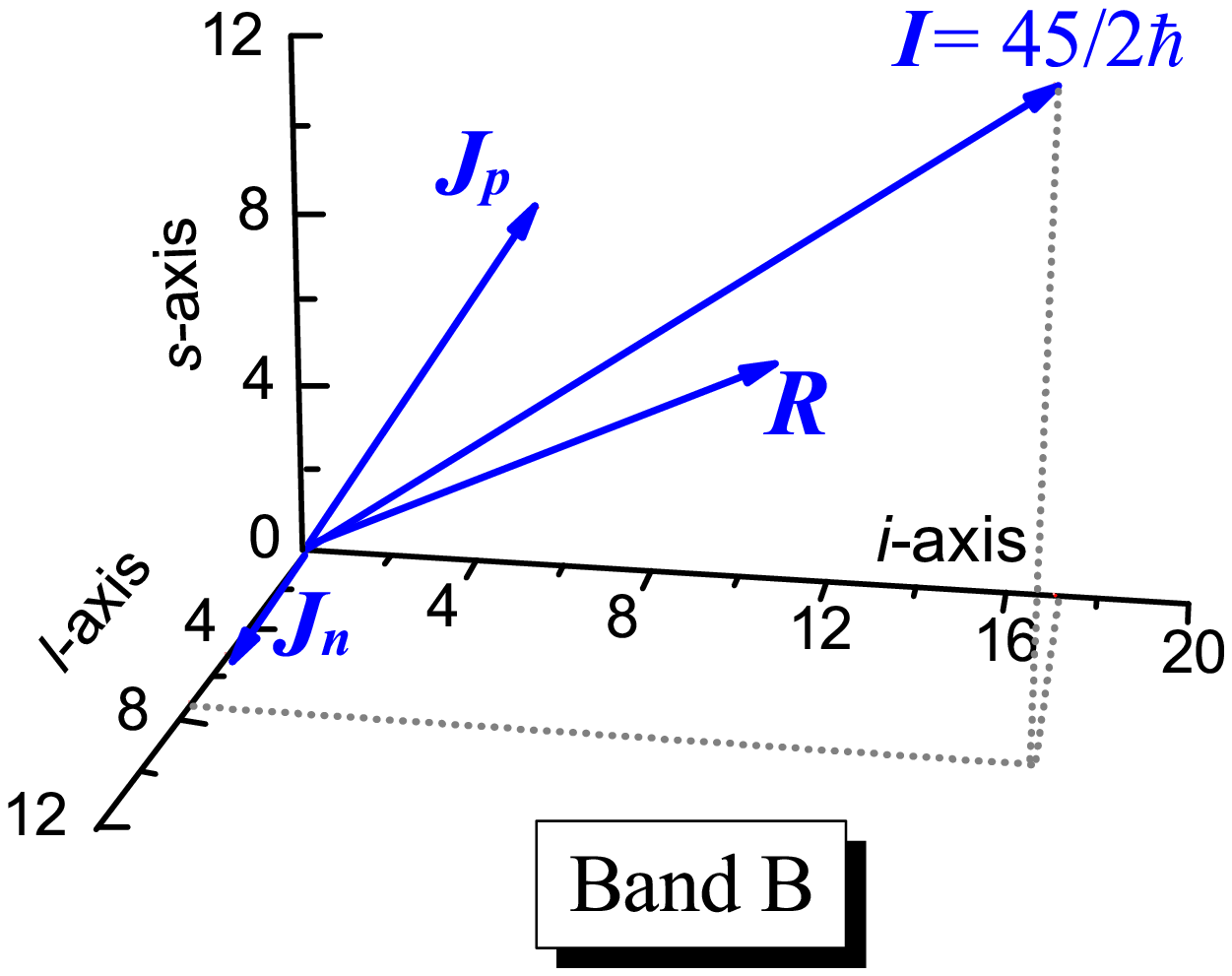}
\caption{ (color online) The angular momentum geometry in PRM for
the doublet bands in $^{135}$Nd at spins $I=$29/2, 39/2, and 45/2. }
\label{fig:spin2}
\end{figure}

The success in reproducing the energy spectra and transition
probabilities for the doublet bands A and B in $^{135}$Nd suggests
that the PRM calculation must correctly account for the structure of
the states. To exhibit their chiral geometry~\cite{Zhang07}, we
calculated for the  bands A and B in $^{135}$Nd the expectation
values of the squared angular momentum components  for the total
nucleus ${I_k} = \sqrt{\langle \hat{I}_{k}^2 \rangle}$, the core
${R_k} = \sqrt{\langle \hat{R}_{k}^2 \rangle}$, the valence neutron
$J_{nk} = \sqrt{\langle \hat{j}_{nk}^2\rangle}$, and the valence
protons
$J_{pk}=\sqrt{\langle(\hat{j}_{(p1)k}+\hat{j}_{(p2)k})^2\rangle}$,
$(k=1, 2, 3)$, which are presented in Fig.~\ref{fig:spin1} and
\ref{fig:spin2}.

As shown in Fig.~\ref{fig:spin1}, for both bands A and B, the
collective core angular momentum  mainly aligns along the
intermediate axis ($i$-axis), because it has the largest moment of
inertia. For $\gamma=22.4^{\circ}$ the ratios between the moments of
inertia are ${\cal J}_i: {\cal J}_s: {\cal J}_l =6.8: 2.6: 1.0 $.
The angular momentum of the $h_{11/2}$ valence neutron hole mainly
aligns along the long axis ($l$-axis) and that of the two $h_{11/2}$
valence protons mainly along the short axis ($s$-axis), which
correspond to the orientation  preferred by their interaction with
the triaxial core \cite{FM97}. To be more precise, ${\bf J_{n}}
\sim5\hbar$ along $l$-axis, and both $\bf{J_p}$ $\sim10\hbar$ and
$\bf{R}$ $\sim6-12\hbar$ lie in the plane spanned by $i$- and $s$-
axis, which together form the chiral geometry of aplanar rotation.

In order to show the picture more clearly and examine the evolution
of the chiral rotation, the root mean square values of the angular
momentum components for the nucleus, the core, the valence neutron,
and protons at spin $I=29/2$, 39/2, and 45/2 are illustrated in
Fig.~\ref{fig:spin2}. The $i$- and $s$- components of $\bf{J_n}$,
and the $l$- components  of ${\bf J_p}$ and ${\bf R}$, which are
negligible, have been ignored for clarity. As the total angular
momentum increases, $\bf{R}$ increases gradually, $\bf{J_n}$ remains
almost unchanged, while $\bf{J_p}$ moves gradually toward the
$i$-axis. The difference between proton and neutron is due to
alignment effect by the Coriolis force, which is much weaker for the
single neutron hole. At spin $I=39/2$, where the doublet bands have
smallest energy difference, the orientations of $\bf{R}$, $\bf{J_n}$
and $\bf{J_p}$ for band A and band B are nearly identical. Hence
around this spin the structure comes closest to the ideal chiral
picture of a left- and a right-handed configurations with equal
components $\bf{R}$, $\bf{J_n}$, and $\bf{J_p}$ along the respective
$i$-, $l$-, and $s$-axes, which weakly communicate.

\begin{figure}[h!]
 \centering
  \includegraphics[width=13cm, bb=0 230 710 800]{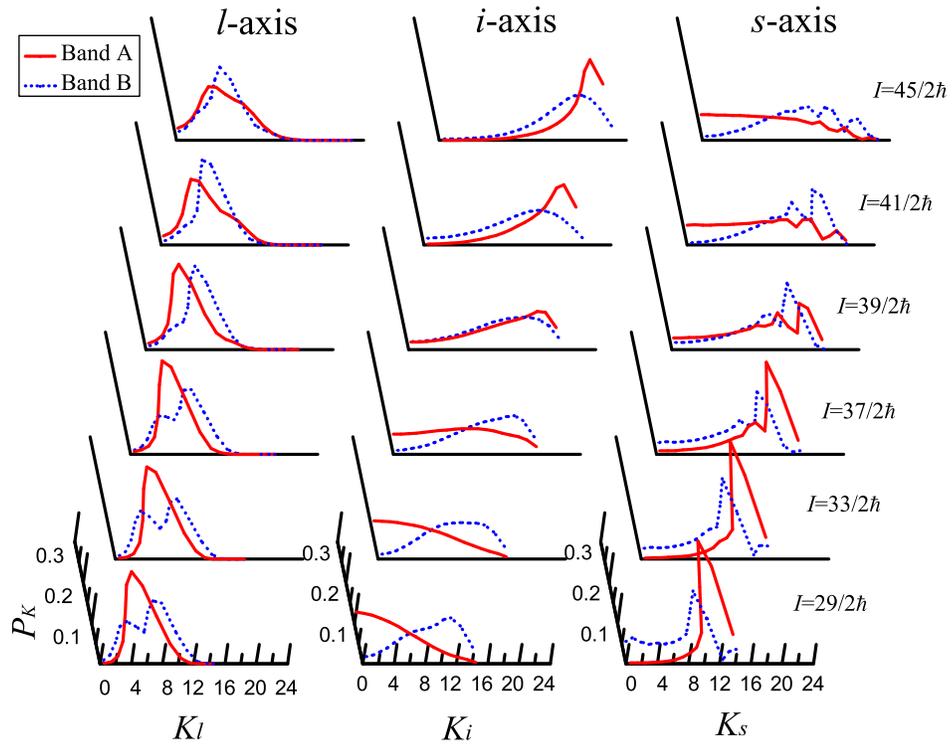}
  \caption{ (color online) The probability distributions for projection of
total angular momentum on the long ($l$-), intermediate ($i$-) and
short ($s$-) axis in PRM for the doublet bands in $^{135}$Nd. }
\label{fig:spin3}
\end{figure}

Further insight into the development of chirality with increasing
angular momentum provides Fig.~\ref{fig:spin3}, which shows the
probability distributions for the projection of the total angular
momentum along the $l$-, $i$- and $s$-axes. For the PRM state
Eq.~(\ref{eq:expansion}), the probability for the projection $K$ of
total angular momentum on quantization axis is
 \beq
   \label{eq:projection} P_{K}= \sum _{\varphi}|c_{K\varphi}|^2.
 \eeq
For triaxiality parameter $\gamma=22.4^\circ$, the $l$-axis is used
for quantization. The distributions with respect to the other axes
are obtained by placing $\gamma$ into another sector, such that the
shape is the same but the principal axes are exchanged.  The
probability distribution of the angular momentum projection on the
$i$-axis is given by Eq.~(\ref{eq:projection}) and
$\gamma=97.6^\circ$; and the distribution of the projection on the
$s$-axis is obtained with $142.4^\circ$.

For spin $I=29/2$, near the band head, the probability distribution
of two bands differ as expected for a chiral vibration. For the
lower band A the maximum probability for the $i$-axis appears at
$K_i=0$, whereas the probability for the higher band B is zero at
$K_i=0$, having its peak at $K_i=25/2$. These are the typical
probability distributions expected for the zero phonon state (A),
which has a wave function that is symmetric in the projection on the
$i$-axis, and the one phonon state (B), which is antisymmetric in
the same degree of freedom. The probability distributions with
respect to the $l$-axis have a peak near $K_l={11/2}$, which is
generated by the $h_{11/2}$ neutron hole. The probability
distributions with respect to the $s$-axis have a peak near $K_s=8$,
which is generated by the pair of $h_{11/2}$ protons. Hence the
chiral vibration consists in an oscillation of the collective
angular momentum vector $\bf{R}$ through the $s$-$l$-plane. This
reveals the structure of the chiral vibration calculated by means of
TAC+RPA in Ref.~\cite{Mukhopadhyay07}.

At spin $I=39/2$, the probability distributions for band A and B are
very similar. The distributions are peaked at
$K_l=7,~K_i=17,K_s=17$, which corresponds to a vector with an
orientation close to the vector ${\bf I}$ in Fig.~\ref{fig:spin2}.
The distributions show the characteristics of static chirality. The
finite values of $P(K_i=0)$ and $P(K_s=0)$ reflect the tunneling
between the left- and right-handed configuration, which is
responsible for the remaining 76 keV energy difference between the
bands. The tunneling is two-dimensional: along the $i$- and
$s$-axes. If it was only along the $i$-axis, band B should have
$P(K_i=0)=0$, as for the chiral vibration. The well developed
tunneling regime is restricted to $I=39/2$. For higher spin, where
the energy difference between the chiral partners increases, they
attain vibration character again. This is reflected by the
increasing differences between the probability distributions of
bands A and B. In particular the fact that $P(K_s=0)$ is finite for
band A (zero phonon) and zero for band B (one phonon) shows that the
motion contains a vibration of the vector ${\bf I}$ through the
$l$-$i$-plane. The two vibrational regimes of our quantal
calculation agree with the ones found by means of the TAC+RPA
calculations in Ref. \cite{Almehed08}.

In summary, we have developed a particle-rotor model, which couples
a triaxial rotor with many valence protons and many valence
neutrons. We used it to investigate the chirality of the
configuration $\pi h_{11/2}^2\otimes\nu h^{-1}_{11/2}$ in the
odd-$A$ nucleus $^{135}$Nd. The energy spectra of the doublet bands,
the reduced probabilities  $B(M1)$ and $B(E2)$ for in-band as well
as for  the interband transitions are reproduced excellently.
Remarkably, the agreement with the data, in particular the energy
splitting, is considerably better than in the case of odd-odd
nuclei~\cite{PengJ03}, indicating that the longer angular momentum
vector of the two protons leads to more pronounced chirality. The
chirality is shown to be a transient phenomenon. The chiral partner
bands start as a soft vibration of the angular momentum
perpendicular to the plane spanned by the short and long axes, where
band A is the zero- and band B the one-phonon state. With increasing
angular momentum the vibration becomes strongly anharmonic,
progressively localizing in  left- and right-handed configurations.
Maximal chirality is reached at $I=39/2$, where the two bands
approach each other closest. At this spin they have very similar
distributions of the angular momenta of the core, the valence
neutron, and the valence  protons in the left- and right-handed
sectors, which reflect reduced left-right tunneling. With further
increasing spin the two bands again develop into the zero and
one-phonon states of a chiral vibration of the angular momentum
about the intermediate axis. The success of the present model
encourages further application for chiral rotation in other nuclei
including the even-even nuclei and settling  the dispute on the
identification and interpretation of chiral doublet bands based on
PRM with $1$-particle-$1$-hole configuration.

\section*{Acknowledgements}
This work is partly supported by Major State Basic Research
Developing Program 2007CB815000, the National Natural Science
Foundation of China under Grant Nos. 10875074, 10775004, 10705004,
10505002, 10435010 and 10221003 as well as by the US Department of
Energy grant DE-FG02-95ER4093.


\end{document}